\documentclass[prl,aps,twocolumn,showpacs]{revtex4}

\usepackage{graphicx}
\graphicspath{{Figures/}}
\DeclareGraphicsExtensions{.eps,.ps}
\usepackage{amsmath}
\usepackage{amssymb}

\begin{document}

\title{Theory of metastability in simple metal nanowires}

\author{J.~B\"urki, C.~A.~Stafford, D.~L.~Stein}

\affiliation{Department of Physics, University of Arizona,
1118 E.\ Fourth Street, Tucson, AZ 85721}

\date{\today}

\begin{abstract}
Thermally induced conductance jumps of metal nanowires are modeled using
stochastic Ginzburg-Landau field theories.  Changes in radius are
predicted to occur via the nucleation of surface kinks at the wire ends,
consistent with recent electron microscopy studies. The activation rate
displays nontrivial dependence on nanowire length, and undergoes first- or
second-order-like transitions as a function of length.  The activation
barriers of the most stable structures are predicted to be {\em universal}, 
i.e., independent of the radius of the wire, and proportional
to the square root of the surface tension. The reduction of the activation
barrier under strain is also determined.
\end{abstract}

\pacs{
%%%%%%%   PACS numbers   %%%%%%%%%%%%%%%%%%%%%%%%%%%%%%%%%%%%%%%%%%%%%%%%%%%%%%%%%%%%%
05.40.-a, % Fluctuation phenomena, random processes, noise, and Brownian motion 
68.65.La  % Quantum wires: structure and nonelectronic properties
}
%%%%%%%  end of PACS numbers  %%%%%%%%%%%%%%%%%%%%%%%%%%%%%%%%%%%%%%%%%%%%%%%%%%%%%%%%

\maketitle\vskip2pc

Metal nanowires have attracted considerable interest in the past decade due to their remarkable
transport and structural properties \cite{Agrait03}.
Long gold and silver nanowires were observed to form spontaneously under electron irradiation
\cite{Kondo97,Rodrigues02b,Oshima03},
and appear to be surprisingly stable; even the thinnest gold wires, essentially
a chain of atoms, have lifetimes of the order of seconds at room temperature \cite{Smit03a}.
Nanowires formed from alkali metals are significantly less long-lived, but 
exhibit striking correlations between their stability and electrical conductance \cite{Yanson99,Yanson01a}.
That these filamentary structures are stable at all is rather counterintuitive \cite{Kassubek01,Zhang03}, but 
can be explained by electron-shell effects \cite{Yanson99,Kassubek01,Zhang03,Burki03}.
Nonetheless, these nanostructures are only metastable, and understanding their lifetimes is of fundamental interest
both for their potential applications in nanoelectronics and as an interesting problem in nanoscale nonlinear dynamics.

In this Letter, we introduce a continuum approach to study the lifetimes
of monovalent metal nanowires.  Our starting point is the nanoscale
free-electron model~\cite{stafford97a}, in which the ionic medium is
treated as an incompressible continuum, and electron-confinement effects are
treated exactly, or through a semiclassical approximation~\cite{Kassubek01,Zhang03,Burki03}.
This approach is most appropriate \cite{Zhang03}
for studying simple metals, whose properties are determined largely by the conduction-band $s$-electrons, 
and for nanowires of `intermediate' thickness: thin enough so that electron-shell effects dominate the energetics, 
but not so thin that a continuum approach is unjustified.  The inclusion of thermal fluctuations is modeled using a
stochastic Ginzburg-Landau classical field theory, which provides
a self-consistent description of the fluctuation-induced thinning/growth
of nanowires.

Our theory provides
quantitative estimates of the lifetimes for alkali nanowires with
electrical conductance~$G$ in the range~$3\leqslant G/G_0\leqslant100$, where
$G_0=2e^2/h$~is the conductance quantum.  In addition, we predict a
universality of the typical escape barrier for a given metal, independent
of the wire radius, with a value proportional to~$\sqrt{\sigma}$, where~$\sigma$
is the surface tension of the material.  Our model can therefore account
qualitatively for the large difference in the observed stability of alkali
vs.~noble metal nanowires.  It also predicts a sharp decrease of the escape
barrier under strain.

\begin{figure}[b]
  \includegraphics[angle=0,width=1\columnwidth]{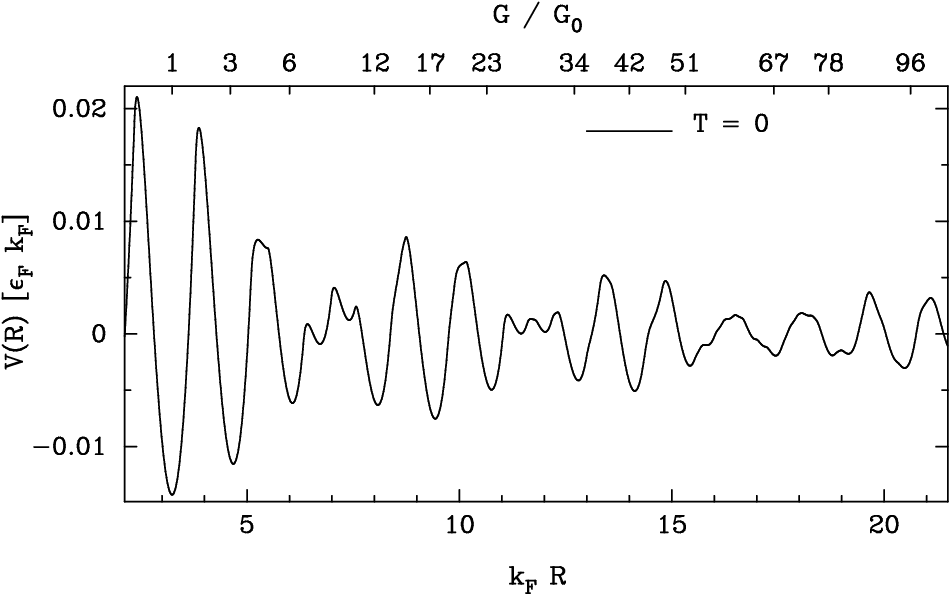}
  \caption{Electron-shell potential $V(R)$ at zero temperature.
  	The top axis shows the conductance values of the most stable wires in units of the conductance quantum, $G_0=2e^2/h$.
	\label{fig:potential}}
\end{figure}

We consider a cylindrical wire suspended between two metallic electrodes, with which it can exchange
both atoms and electrons.  The resulting energetics
is described through an ionic grand canonical potential
\begin{equation}\label{eq:omega_a}
  \Omega_a = \Omega_e-\mu_a{\cal N}_a,
\end{equation}
where $\Omega_e$ is the free energy for a fixed number ${\cal N}_a$ of atoms
and $\mu_a$ is the chemical potential for a surface atom in the electrodes.  In the
Born-Oppenheimer approximation, $\Omega_e$ is just the
electronic grand-canonical potential, 
and can be written as a Weyl expansion plus an electron-shell correction \cite{Burki03}
\begin{equation}\label{eq:omega_e}
  \Omega_e = -\omega{\cal V}+\sigma{\cal S}
+ \int_0^L dz \, V(R(z)),
\end{equation}
where ${\cal V}$, ${\cal S}$, and $L$ are, respectively, the volume, surface area, and length of the wire, 
$\omega$ and $\sigma$ are material-dependent coefficients, and
$V(R)$, shown in Fig.~\ref{fig:potential}, is a mesoscopic electron-shell potential \cite{Burki03} 
that describes electronic quantum-size effects.

In the presence of thermal noise, a wire's radius will fluctuate as a
function of time $t$ and position $z$ measured along the wire's axis:
$R(z,t)=\bar{R}+\phi(z,t)$ for a wire of radius $\bar{R}$ at zero temperature. 
The wire energy~(\ref{eq:omega_a}) may be expanded as
\begin{equation}\label{eq:omega_cyl}
  \Omega_a[\bar{R}, \phi] = \Omega_a(\bar{R}) + {\cal H}[\phi],
\end{equation}
where ${\cal H}[\phi]$ is the energy of the
fluctuations.  Keeping only the lowest-order terms in $\partial_z\phi$, one finds
\begin{equation}\label{eq:H}
  {\cal H}[\phi] =
     \int_0^L\!\text{d}z\left[\frac{\kappa}{2}(\partial_z\phi)^2+U(\phi)\right],
\end{equation}
where $\kappa=2\pi\sigma \bar{R}$ and $U(\phi)$ is an effective potential.

A (meta)stable nanowire is in a state of diffusive equilibrium:
\begin{equation}
\label{eq:mu_cyl}
\mu_a = \mu_{\rm cyl}(\bar{R}) = {\cal V}_a \left(\frac{\sigma}{\bar{R}} - \omega + \frac{1}{2\pi \bar{R}} 
\frac{d V(\bar{R})}{d \bar{R}}\right),
\end{equation}
where $\mu_{\rm cyl}$ is the chemical potential of a surface atom in a cylindrical wire \cite{Burki03}
and ${\cal V}_a$ is the volume of an atom.  
Using Eq.\ (\ref{eq:mu_cyl}) in Eq.\ (\ref{eq:omega_a}), one finds
the effective potential
\begin{equation}\label{eq:Vofphi}
  U(\phi) = V(\bar{R}+\phi)-V(\bar{R})-\frac{\pi\sigma}{\bar{R}}\phi^2
- \left(\phi + \frac{\phi^2}{2 \bar{R}} \right) \frac{d V}{d \bar{R}}.
\end{equation}
A stable wire must satisfy $U^\prime(0)=0$ and $U^{\prime\prime}(0)>0$.  The first condition is
satisfied automatically.
The second condition is equivalent to the
requirement $d \mu_{\rm cyl}/d\bar{R} > 0$, and was previously used to determine 
the linear stability of metal nanowires \cite{Zhang03}.
The most stable wires correspond to the minima of $V(R)$ (cf.\ Fig.\ \ref{fig:potential}); however, the stable zones
span finite intervals of radius about the minima \cite{Zhang03}.

The radius fluctuations~$\phi(z,t)$ due to thermal noise
can be treated as a classical field on $[0,L]$,
with dynamics governed by the stochastic Ginzburg-Landau (GL) equation 
\begin{equation}
\label{eq:GL}
\frac{\partial \phi(z,t)}{\partial t} = \kappa \frac{\partial^2 \phi}{\partial z^2} -
\frac{\partial U}{\partial\phi} + (2T)^{1/2}\xi(z,t),
\end{equation}
where $\xi(z,t)$ is unit-strength spatiotemporal white noise.
The zero-noise dynamics is {\em gradient}, that is,
$\dot{\phi} = - \delta {\cal H}/\delta \phi$ at zero temperature.
In~(\ref{eq:GL}), time is measured in units of a microscopic timescale
describing the short-wavelength cutoff of the surface
dynamics~\cite{Zhang03,Smit03a} which is of order the inverse Debye frequency
$\nu_D^{-1}$.

At nonzero temperature, thermal fluctuations can 
drive a nanowire to escape from the metastable configuration $\phi=0$, leading to a finite lifetime of such a nanostructure.
The escape process occurs via nucleation of a ``droplet'' of one stable configuration in the background of the other,
subsequently quickly spreading to fill the entire spatial domain.
A transition from one
metastable state to another must proceed via a pathway of states, accessed through random thermal fluctuations,
that first goes ``uphill'' in energy from the starting configuration.  Because these fluctuations are
exponentially suppressed as their energy increases, there is at low
temperature a preferred transition configuration (saddle) that lies between adjacent minima.  
The activation rate is given in the $T\to0$ limit by the Kramers formula~\cite{HTB90}
\begin{equation}
\label{eq:Kramers}
\Gamma\sim \Gamma_0\exp(-\Delta E/T)\, .
\end{equation}
Here $\Delta E$ is the activation barrier, the difference in energy between
the saddle and the starting metastable configuration, and $\Gamma_0$ is the
rate prefactor.

The quantities $\Delta E$ and $\Gamma_0$ depend on the microscopic
parameters of the nanowire through $\kappa$ and the details of the potential (\ref{eq:Vofphi}),
on the length~$L$ of the wire, and on the
choice of boundary conditions at the endpoints $z=0$ and~$z=L$.
Simulations of the structural dynamics under surface
self-diffusion~\cite{Burki03} suggest that the connection of the
wire to the electrodes is best described by Neumann boundary conditions,
$\partial_z\phi|_{0,L}=0$.
These boundary conditions force nucleation to
begin at the endpoints, consistent with experimental observations \cite{Oshima03}.

The saddle configurations are time-independent solutions of the
zero-noise GL equation \cite{HTB90}, and can be obtained by numerical integration of Eq.\ (\ref{eq:GL}) at $T=0$.
However, we find that for many of the metastable wires, the effective potential $U(\phi)$ can be approximated locally by a cubic potential
\begin{equation}\label{eq:cubic}
  U^{(\pm)}(\phi)=-\alpha\tilde{\phi}_\pm + \frac{\beta}{3}\big(\tilde{\phi}_\pm\big)^3,
\end{equation}
where $\tilde{\phi}_\pm=\sqrt{\frac{\alpha}{\beta}}\mp\phi$ and $\alpha,\beta > 0$.  
The potential $U^{(-)}$ ($U^{(+)}$) biases fluctuations toward smaller (larger) radii.
It is useful to scale out the various constants in the model by
introducing the dimensionless variables $x=z/L_0$ and $u=(\beta/\alpha)^{1/2}\tilde\phi$,
where $L_0=\kappa^{1/2}/(\alpha\beta)^{1/4}$ and $E_0=\kappa^{1/2}\alpha^{5/4}/\beta^{3/4}$
are characteristic length and energy scales.
The energy functional then becomes
\begin{equation}
\label{eq:asymscaled}
\frac{{\cal H}[u]}{E_0}=\int_{0}^{\ell} \left[\frac{1}{2}\left(u^\prime\right)^2  -u + \frac{1}{3}u^3\right]\,dx,
\end{equation}
where $\ell \equiv L/L_0$ and $u^\prime = \partial u/\partial x$.

Metastable and saddle configurations are stationary functions of Eq.\ (\ref{eq:asymscaled}),
and therefore obey the Euler-Lagrange equation
$u''=-1+u^2$.  With Neumann boundary conditions, the uniform stable state
is the constant state $u_s=+1$, and there exists a uniform unstable state
$u_u=-1$.  We will see that the latter is the saddle for
$\ell<\ell_c=\pi/\sqrt{2}$.  At $\ell_c$ a transition occurs~\cite{MS01},
and above it the saddle is nonuniform.  It consists of an `instanton' localized at one end of the wire, and
is given by~\cite{St04}
\begin{equation}\label{eq:instanton}
  u(x) = \frac{2-m}{\sqrt{\xi(m)}}
    - \frac{3}{\sqrt{\xi(m)}}
    {\rm dn}^2\left(\frac{x}{\sqrt{2}\xi(m)^{1/4}}\Big| m\right)\, ,
\end{equation}
where ${\rm dn}(\cdot\mid m)$ is the Jacobi elliptic $\rm dn$ function with
parameter~$m$, with $0\leqslant m\leqslant 1$, and $\xi(m)=m^2-m+1$.  Its half-period is given by~${\bf K}(m)$, the 
complete elliptic integral of the first
kind, a monotonically increasing function of~$m$.
Eq.~(\ref{eq:instanton}) satisfies the Neumann boundary condition when
\begin{equation}\label{eq:asybc}
  \ell = \sqrt{2}\xi(m)^{1/4}{\bf K}(m)\, ,
\end{equation}
which (taking $m\to 0^+$) leads to $\ell_c=\pi/\sqrt{2}$.  As
$\ell\to\ell_c^+$, ${\rm dn}(x|0)\to 1$, and the nonuniform saddle reduces
to the uniform state $u_u=-1$.  This is the saddle for all $\ell<\ell_c$.

\begin{figure}[b]
   \includegraphics[width=0.99\columnwidth]{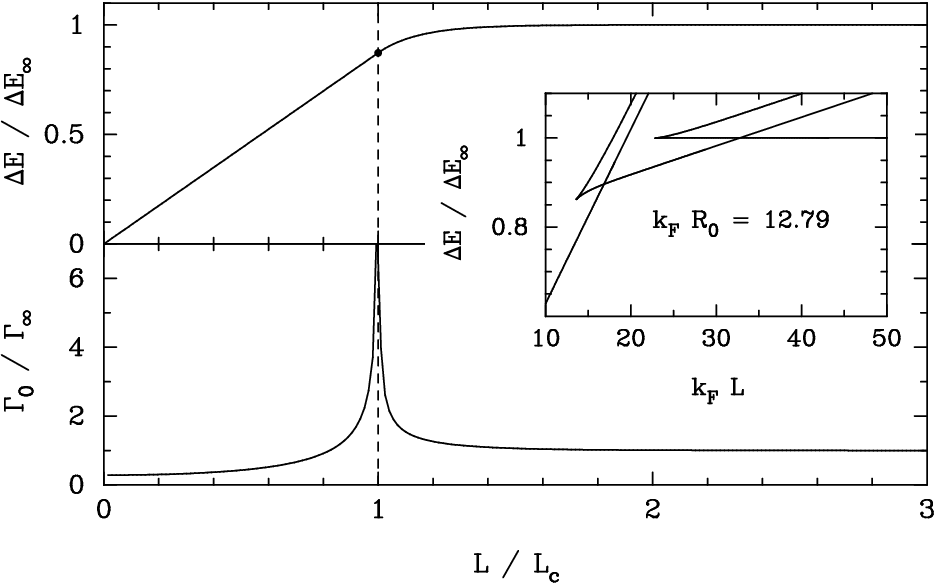}
   \caption[The activation energy.]{The activation energy $\Delta E$ as a
   	function of the wire length~$L$, for the cubic potential with
	Neumann boundary conditions (top).  The dashed line %bullet 
	indicates the critical
	wire length $L_c$ at which the transition takes place.
	The bottom panel shows the prefactor $\Gamma_0$, and the inset
	displays the activation barrier for the full
	potential $U(\phi)$ for $k_FR_0=12.79$, exhibiting a succession of first
	order phase transitions.
	\label{fig:E+G}}
\end{figure}

When the saddle is constant ($\ell\leqslant\ell_c$), %this
the activation barrier
scales linearly with (reduced) length $\ell$: $\Delta E/E_0 = (4/3)\ell$.
Above $\ell_c$, it is expressible in terms of the complete elliptic
integrals of the first kind ${\bf K}(m)$ and the second kind ${\bf E}(m)$:
\begin{multline}\label{eq:barrier}
  \frac{\Delta E}{E_0} =
    \left[{2-3m-3m^2+2m^3\over 3\xi(m)^{3/2}}+{2\over 3}\right]\ell \\
    + {6\sqrt{2}\over 5\xi(m)^{1/4}}
      \left[2{\bf E}(m)-{(2-m)(1-m)\over\xi(m)}{\bf K}(m)\right]\, .
\end{multline}
As $\ell\to\infty$ (corresponding to $m\to 1^-$),
$\Delta E/E_0\to 12\sqrt{2}/5$.  More generally, we denote the asymptotic value $\lim_{L \to \infty}\Delta E(L) \equiv \Delta E_\infty$.
The activation barrier for the entire range of $\ell$ is shown in Fig.~\ref{fig:E+G}.

Calculation of the prefactor $\Gamma_0$ in the Kramers transition rate
formula is a much more involved matter.  It generally requires an analysis
of the {\em transverse fluctuations\/} about the extremal solutions.  The
general method for determining~$\Gamma_0$ has been discussed
elsewhere~\cite{MS01,St04}; here, we just present results (in units of
the Debye frequency $\nu_D$).
For $\ell<\ell_c$, we find
\begin{equation}\label{eq:g0-}
  \Gamma_0^< = {1\over\pi}{\sinh(\ell\sqrt{2})\over\sin(\ell\sqrt{2})}\, ,
\end{equation}
which diverges as $\ell\to{\ell_c}^-$, with a critical exponent of 1/2.
The divergence arises from a {\it soft mode\/}; one of the eigenmodes
corresponding to small fluctuations about the saddle has vanishing
eigenvalue at $\ell_c$.  This divergence, and its meaning, are discussed
in detail in~\cite{St04}.

For $\ell>\ell_c$, the prefactor is
\begin{multline}\label{eq:g0+}
  \Gamma_0^> = {2-m+ 2\sqrt{4m^2-m+1}\over4\pi\xi(m)^{3/8}}\\
    \times
  \sqrt{(1-m)\sinh\Bigl[2\xi(m)^{1/4}{\bf K}(m)\Bigr]
    \over\xi(m){\bf E}(m)-\frac{1}{2}(1-m)(2-m){\bf K}(m)}.
\end{multline}
This also exhibits a divergence with
a critical exponent of $1/2$ as $\ell\to{\ell_c}^+$.  The prefactor over
the entire range of $\ell$ is shown in Fig.~\ref{fig:E+G}.

The second-order-like transition in activation behavior exhibited in
Fig.~\ref{fig:E+G} is interesting, but generally holds only for transitions
where the potential $U(\phi)$ can be locally approximated
by a smooth potential of quartic or lower order \cite{St04}.  For some of the minima
of~Fig.~\ref{fig:potential}, this is not the case, and the
wire instead exhibits one or more first-order-like transitions \cite{Chudnovsky92}, as
shown in the inset of Fig.~\ref{fig:E+G}.

\begin{figure}[b]
   \includegraphics[width=1\columnwidth]{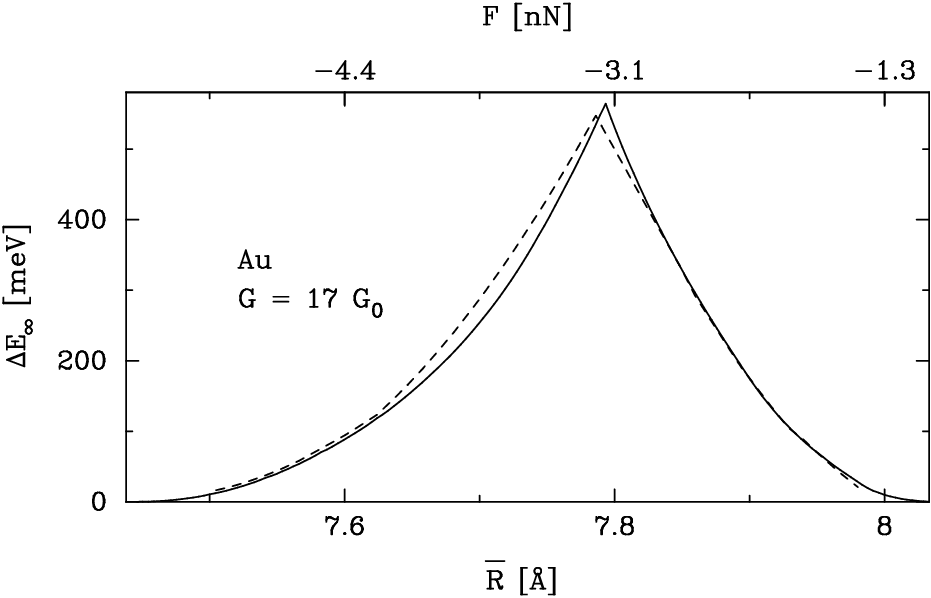}
   \caption[Strained wire.]{The activation energy $\Delta E_\infty$
	as a function of radius for a typical stable zone in Au.
	Solid curve: numerical result for the full potential $U(\phi)$, Eq.\ (\ref{eq:Vofphi}); 
	dashed curve: result from Eq.\ (\ref{eq:barrier}) using the best cubic-polynomial fit to $U(\phi)$.
   	The wire radius is related to the tensile stress (upper axis).
	\label{fig:strain}}
\end{figure}

Figure \ref{fig:strain} shows the activation barrier $\Delta E_\infty$
as a function of radius $\bar{R}$ 
for a typical metastable wire, corresponding to the conductance plateau at $G=17\,G_0$ in Au.
Very good agreement is found between
the numerical result for the full potential (\ref{eq:Vofphi}) (solid curve) and the result from Eq.\ (\ref{eq:barrier})
using the best-fit cubic polynomial $U^{(\pm)}$ (dashed curve).
Under strain, $\bar{R}$ varies elastically;
the corresponding stress in the wire is shown on the upper axis.
A stress of a fraction of a nanonewton can significantly change the activation barrier, and even change the {\em direction} of escape. 
The maximum value of $\Delta E_\infty$
occurs at the cusp, where the activation barriers for {\em thinning} and {\em growth} are equal.

\begin{table}[t]
  \caption{The lifetime $\tau$ (in seconds) for various cylindrical sodium
    nanowires at temperatures from 75K to 125K.
    Here $G$ is the electrical conductance of the wire,
    $L_c$ is the critical length above which the lifetime may
    be approximated by $\tau \approx \nu_D^{-1} \exp(\Delta E_\infty/T)$, and
    $\Delta E_\infty$ is the activation energy for an infinitely long wire.
    Note that wires shorter than $L_c$ are predicted to have shorter lifetimes.
    \label{tab:lifetime}}
  \begin{center}
    \begin{tabular}{|c||c|c|c|c|c|}
      \hline
      \rule[-1.5ex]{0pt}{4.5ex} $G$ & $L_c $ & $\Delta E_\infty $ &
      \multicolumn{3}{c|}{$\tau$ [s]} \\
      \cline{4-6}\rule[-1.5ex]{0pt}{4.5ex}
      [$G_0$] & [\AA] & [meV] & $75\;$K & $100\;$K & $125\;$K \\
      \hline\hline
      \rule[0ex]{0pt}{3.ex}%
       % Values from cubic fit:
       3  &  2.8 & 250 & $4\times10^5$ & 2                & $5\times10^{-3}$ \\
       6  &  4.3 & 200 & 7             & $3\times10^{-3}$ & $3\times10^{-5}$ \\
       17 &  5.0 & 260 & $7\times10^5$ & 3                & $8\times10^{-3}$ \\
       23 &  6.1 & 230 & $2\times10^3$ & 0.2              & $9\times10^{-4}$ \\
       42 &  7.2 & 250 & $2\times10^5$ & 1                & $10^{-3}$ \\
       51 &  6.8 & 190 & 1	       & $8\times10^{-4}$ & $10^{-4}$        \\
       67 & 18.8 & 180 & 0.6           & $5\times10^{-4}$ & $7\times10^{-6}$ \\
      \rule[-1.5ex]{0pt}{3.ex}%
       96 & 11.4 & 250 & $10^5$	       & 0.8	          & $3\times10^{-3}$ \\
      \hline
    \end{tabular}
  \end{center}
\end{table}

The most stable structures, corresponding to the maximum values of $\Delta E_\infty$, occur at (or near) the minima of the electron-shell
potential, $V(R)$ (Fig.\ \ref{fig:potential}).
The lifetimes of these equilibrated structures are limited by thinning, since the total energy of the wire is lowered by reducing its volume.
We thus fit the effective potential at these minima to the form $U^{(-)}$. 
Table~\ref{tab:lifetime} lists critical lengths $L_c$, activation barriers $\Delta E_\infty$,
and lifetimes $\tau=1/\Gamma$, Eq.~(\ref{eq:Kramers}), at various temperatures for
Na nanowires.  (Only the minima that are well-fit by $U^{(-)}$ are shown.)
The temperature dependence of $\tau$ shows that the lifetime of Na nanowires drops below the threshold for observation in 
break-junction experiments as the temperature is increased from 75K to 125K.  This behavior can explain the observed temperature 
dependence of conductance histograms for Na nanowires \cite{Yanson99}, which show clear peaks at conductances near the predicted values
at temperatures below 100K, but were not reported at higher temperatures.
The increase of $L_c$ with $G$, shown in Table \ref{tab:lifetime}, may also explain the observed exponential decrease in the heights of
the conductance peaks with increasing conductance \cite{Yanson99}, since the thicker contacts are more likely to be shorter than $L_c$,
and hence to have exponentially reduced lifetimes.

An important prediction given in Table \ref{tab:lifetime} is that the
lifetimes of the most stable nanowires, while they do exhibit significant
variations from one conductance plateau to another, do not vary systematically
as a function of radius; the activation barriers in Table \ref{tab:lifetime}
vary by only about 30\% from one plateau to another, and the wire with a conductance of
$96 G_0$ has essentially the same lifetime as that with a conductance of
$3 G_0$.  In this sense, the activation barrier is found to be {\em universal}:  in any conductance interval, there are very short-lived wires
(not shown in Table \ref{tab:lifetime}) with very small activation barriers, while the longest-lived wires have activation barriers of a universal
size
\begin{equation}\label{eq:Delta_E}
  \Delta E_\infty\, \simeq \, 0.6
    \left(\frac{\hbar^2 \sigma}{m_e}\right)^{1/2}\!\!\!\!\!,
\end{equation}
depending only on the surface tension of the material.
Here $m_e$ is the conduction-band effective mass, which is comparable to the free-electron rest mass.
The fact that the typical activation energy (\ref{eq:Delta_E}) is independent of $\bar{R}$ is a consequence of the virial theorem:
Since the instanton is a stationary state of Eq.\ (\ref{eq:H}), the bending energy $\langle \frac{\kappa}{2}
(\partial_z \phi)^2\rangle$ is proportional to $\langle U(\phi)\rangle$.  Since $\kappa \sim \sigma \bar{R}$ and $V \sim 1/\bar{R}$
\cite{Burki03}, 
this implies that
the characteristic size of the instanton $L_c \sim \sqrt{\sigma} \bar{R}$ and $\Delta E_\infty \sim \sqrt{\sigma}$.

\begin{table}[t]
  \caption{Fermi energy, surface tension, and typical critical length and activation
    barrier for various alkali and noble metals. The surface tension values
    are extrapolations to zero temperature from Ref.\ \cite{Tyson77}.
    \label{tab:materials}}
  \begin{center}
    \begin{tabular}{|c||c|c|c|c|c||c|c|c|}
	\hline
	Metal & Li & Na & K & Rb & Cs & Cu & Ag & Au \\
	\hline\hline
	$\varepsilon_F$[eV] & 4.74 & 3.24 & 2.12 & 1.85 & 1.59 & 7.00 & 5.49 & 5.53\\
        $\sigma$ [N/m] & 0.52 & 0.26 & 0.14 & 0.12 & 0.09 & 1.78 & 1.24 & 1.50 \\[1mm]
	\hline
	$L_c/\bar{R}$ & 0.67 & 0.71 & 0.81 & 0.84 & 0.88 & 0.83 & 0.88 & 0.97 \\
	$\Delta E_{\infty}$ [meV] & 290 & 200 & 150 & 140 & 120 & 530 & 440 & 490
	\\[1mm]
	\hline
    \end{tabular}
  \end{center}
\end{table}

Table~\ref{tab:materials} lists typical activation barriers
and critical lengths for various alkali and noble metals. 
It shows that noble metal nanowires should have much longer lifetimes than alkali metal nanowires, due to their 
larger surface tension coefficients.
This prediction is consistent with experimental observations \cite{Kondo97,Rodrigues02b,Oshima03,Smit03a,Yanson99,Yanson01a}, 
although our estimated activation barriers for noble metal nanowires are still too small to account for their observed stability at 
room temperature.
This discrepancy may stem from the neglect of $d$-electrons in our model (except 
inasmuch as they enhance $\sigma$ compared to the
free-electron value), or due to the presence of impurities which passivate the surface, thereby raising the activation barrier above its
intrinsic value.

\begin{acknowledgments}
J.B. and C.A.S. acknowledge support from NSF Grant No.\ DMR0312028. D.L.S. acknowledges support from NSF Grant Nos.\ PHY0099484 and PHY0351964.
\end{acknowledgments}

\bibliography{refs}

\end{document}